\documentclass[aps,pra,superscriptaddress,twocolumn,10pt,a4paper]{revtex4-1}
\usepackage{graphicx}
\usepackage{braket}
\pdfoutput=1

\begin{document}

\title[Quantifying the quantum gate fidelity of single-atom spin qubits in silicon]{Quantifying the quantum gate fidelity of single-atom spin qubits in silicon by randomized benchmarking}

\author{J T Muhonen}
\email{juha.muhonen@unsw.edu.au}
\affiliation{Centre for Quantum Computation and Communication Technology, School of Electrical Engineering and Telecommunications, UNSW Australia, Sydney, New South Wales 2052, Australia}

\author{A Laucht}
\affiliation{Centre for Quantum Computation and Communication Technology, School of Electrical Engineering and Telecommunications, UNSW Australia, Sydney, New South Wales 2052, Australia}

\author{S Simmons}
\affiliation{Centre for Quantum Computation and Communication Technology, School of Electrical Engineering and Telecommunications, UNSW Australia, Sydney, New South Wales 2052, Australia}

\author{J P Dehollain}
\affiliation{Centre for Quantum Computation and Communication Technology, School of Electrical Engineering and Telecommunications, UNSW Australia, Sydney, New South Wales 2052, Australia}

\author{R Kalra}
\affiliation{Centre for Quantum Computation and Communication Technology, School of Electrical Engineering and Telecommunications, UNSW Australia, Sydney, New South Wales 2052, Australia}

\author{F E Hudson}
\affiliation{Centre for Quantum Computation and Communication Technology, School of Electrical Engineering and Telecommunications, UNSW Australia, Sydney, New South Wales 2052, Australia}

\author{S Freer}
\affiliation{Centre for Quantum Computation and Communication Technology, School of Electrical Engineering and Telecommunications, UNSW Australia, Sydney, New South Wales 2052, Australia}

\author{K M Itoh}
\affiliation{School of Fundamental Science and Technology, Keio University, 3-14-1 Hiyoshi, 223-8522, Japan}

\author{D N Jamieson}
\affiliation{Centre for Quantum Computation and Communication Technology, School of Physics, University of Melbourne, Melbourne, Victoria 3010, Australia}

\author{J C McCallum}
\affiliation{Centre for Quantum Computation and Communication Technology, School of Physics, University of Melbourne, Melbourne, Victoria 3010, Australia}

\author{A S Dzurak}
\affiliation{Centre for Quantum Computation and Communication Technology, School of Electrical Engineering and Telecommunications, UNSW Australia, Sydney, New South Wales 2052, Australia}

\author{A Morello}
\email{a.morello@unsw.edu.au}
\affiliation{Centre for Quantum Computation and Communication Technology, School of Electrical Engineering and Telecommunications, UNSW Australia, Sydney, New South Wales 2052, Australia}

\begin{abstract}
Building upon the demonstration of coherent control and single-shot readout of the electron and nuclear spins of individual $^{31}$P atoms in silicon, we present here a systematic experimental estimate of quantum gate fidelities using randomized benchmarking of 1-qubit gates in the Clifford group. We apply this analysis to the electron and the ionized $^{31}$P nucleus of a single P donor in isotopically purified $^{28}$Si. We find average gate fidelities of 99.95\% for the electron, and  99.99\% for the nuclear spin. These values are above certain error correction thresholds, and demonstrate the potential of donor-based quantum computing in silicon. By studying the influence of the shape and power of the control pulses, we find evidence that the present limitation to the gate fidelity is mostly related to the external hardware, and not the intrinsic behaviour of the qubit.
\end{abstract}

\maketitle

\section{Introduction}

The discovery of quantum error correction is among the most important landmarks in quantum information science \cite{Shor1995,Steane1996,Nielsen2000,Raussendorf2012}. Together with the steady improvement in the coherence and fidelity of various physical quantum bit (qubit) platforms, it represents one of the main motivations for the investment in quantum information technologies. In general, uncontrolled interactions between the qubit and its environment can cause loss of coherence and gate errors. Quantum error correcting codes, however, guarantee that the errors can be recovered, provided that the average error rate is below a certain fault-tolerance threshold \cite{Knill1998}. The threshold is strongly dependent on the quantum computer architecture: for instance, a bilinear nearest-neighbour array requires errors below $10^{-6}$ to achieve fault-tolerance \cite{Stephens2008}. More recent ideas have shown the tantalizing prospect of fault-tolerance error correction with error rates that can be as high as 10$^{-2}$ \cite{Knill2005,Fowler2012}. This level of gate fidelity has become accessible to the most advanced qubit platforms, such as superconducting qubits \cite{Barends2014} and trapped ions \cite{Benhelm2008}.

Quantifying gate errors is, however, not trivial. The most commonly used protocol, quantum process tomography (QPT) \cite{Chuang1997}, is based upon preparing different input states (chosen to form a complete basis) and processing them identically many times. The output states are then characterized with quantum state tomography to extract all the components of the process matrix and quantify the process errors. QPT can in principle be applied to any process and state space, but it is not scalable to large number of qubits since the number of measurements required to completely map a multi-qubit process increases exponentially with the number of qubits involved.

In addition, and more importantly for the single qubit system discussed here, quantum process tomography is also sensitive to errors in the state preparation and measurement (SPAM errors) and cannot distinguish between these and pure control errors \cite{Chow2009}. Therefore, with QPT it is only possible to characterize gate fidelities if the gate errors are larger than preparation and readout errors. This is an unsatisfying situation, since fault-tolerant quantum computation normally poses much more stringent requirements on gate fidelities than on initialization and readout.

Randomized benchmarking \cite{Emerson2005,Knill2008} has gained attention in recent years as a scalable solution for determining gate fidelities, and has been experimentally demonstrated with e.g. atomic ions \cite{Olmschenk2010,Brown2011}, nuclear magnetic resonance \cite{Ryan2009} and superconducting qubits \cite{Chow2009,Barends2014}. The goal of randomized benchmarking protocols is to extract the \textit{average gate fidelity}, defined as the average fidelity of the output state over pure input states.  The fidelity of the output state $\mathcal{E}(\rho)$ as compared to the ideal output state $U(\rho)$ is defined as \cite{Nielsen2000}
\begin{equation}
 \mathcal{F}_U = \left(\textrm{tr}\sqrt{\sqrt{\mathcal{E}(\rho)}U(\rho)\sqrt{\mathcal{E}(\rho)}} \right)^2,
 \end{equation}
where $U$ is the ideal gate operation, $\mathcal{E}$ the actual operation and $\rho$ is the density matrix describing the input state.

Here we adopt a benchmarking method based upon the construction and application of random sequences of gates that belong to the Clifford group. It has been shown \cite{Magesan2011,Magesan2012,Epstein2014} that the average fidelity of a Clifford gate can be extracted from the fidelity of the final state of the qubit, averaged over several random sequences of Clifford gates, having started from a fixed input state. This provides a scalable benchmarking method with well-defined conditions of applicability, and which does not depend on SPAM errors \cite{Magesan2011,Magesan2012,Epstein2014}. The Clifford group does not provide a universal set of quantum gates, but a universal set can be constructed from them by the addition of only one gate or, alternatively, by using ancilla qubits and their measurements \cite{Nielsen2000}. Also, the Clifford gates play an important role in the error correction schemes based on stabilizer codes \cite{Gottesman1998}.

Spin qubits based upon the electron or nuclear spin of phosphorus atoms in isotopically purified $^{28}$Si \cite{Ager2005} are known to have extraordinary coherence times \cite{Tyryshkin2012,Saeedi2013}, and recent experiments have shown that such record coherence can be retained also at the single-atom level in functional nanoelectronic devices \cite{Muhonen2014}. Here, we present a thorough investigation of the gate fidelities of the single-atom spin qubit device described in Ref. \cite{Muhonen2014}. We focus on the gate fidelity of qubits represented by the electron (e$^-$) and ionized nuclear spin ($^{31}$P$^+$) with randomized benchmarking protocols using Clifford-group gates. We show that all 1-qubit gate fidelities are consistently above 99.8~\% for the electron (with a deduced average of 99.95~\% from a long sequence of Clifford gates) and above 99.9~\% for the nucleus (with an average of 99.99~\%). These results, combined with the previously reported record coherence times, show that individual dopants in $^{28}$Si are one of the best physical realizations of quantum bits.

\section{Single-atom spin qubit device}

Our qubit system is a single substitutional $^{31}$P donor in silicon, fabricated and operated as described in detail in \cite{Morello2009,Morello2010,Pla2012,Pla2013,Muhonen2014}. In particular, the system used in the present experiment is the same as Device B in \cite{Muhonen2014}. 
Both the donor-bound electron (e$^-$) and the nucleus ($^{31}$P) possess a spin $1/2$ and each encode a single qubit. The qubit logic states are the simple spin up/down eigenstates, which we denote with $\ket{\uparrow}$, $\ket{\downarrow}$ for the electron, and $\ket{\Uparrow}$, $\ket{\Downarrow}$ for the nucleus. The electron and the nucleus are coupled by the contact hyperfine interaction $A$. In the presence of a large magnetic field $B_0$, the resulting eigenstates are, to a very good approximation, the separable tensor products of the electro-nuclear basis states ($\ket{\Uparrow\uparrow}$, $\ket{\Uparrow\downarrow}$, $\ket{\Downarrow\uparrow}$, $\ket{\Downarrow\downarrow}$).

Arbitrary quantum states are encoded on the e$^-$ qubit by applying pulses of oscillating magnetic field $B_1$ at the frequencies corresponding to the electron spin resonance (ESR), $\nu_{e1,2} \approx \gamma_e B_0 \mp A/2$ \cite{Pla2012}, where $\gamma_e = 27.97$~GHz/T, and $A = 96.9$~MHz in this specific device \cite{Muhonen2014}. In the randomized benchmarking experiments discussed below, we operated the e$^-$ qubit at $\nu_{e2} \approx \gamma_e B_0 + A/2$ while the nuclear spin was in the $\ket{\Uparrow}$ state. The nuclear spin qubit was operated in the ionized charge state of the donor (D$^+$), where the nuclear magnetic resonance (NMR) frequency is simply $\nu_{n+} = \gamma_n B_0$, where $\gamma_n = 17.23$~MHz/T is nuclear gyromagnetic ratio \cite{Pla2013}. The experiments were performed in high magnetic fields ($B_0 \approx 1.5$~T applied along the [110] crystal axis of Si) and low temperatures (electron temperature $T_{\rm{el}} \approx 100$~mK). Microwave control fields $B_1$ are produced by a broadband on-chip microwave antenna \cite{Dehollain2013} terminating $\sim 100$~nm away from the donor qubit. At microwave frequencies, the cable and the cold attenuators connecting the source to the on-chip antenna provide $\sim 40$~dB attenuation.

\begin{figure}
\centering
\includegraphics[width=0.45\textwidth]{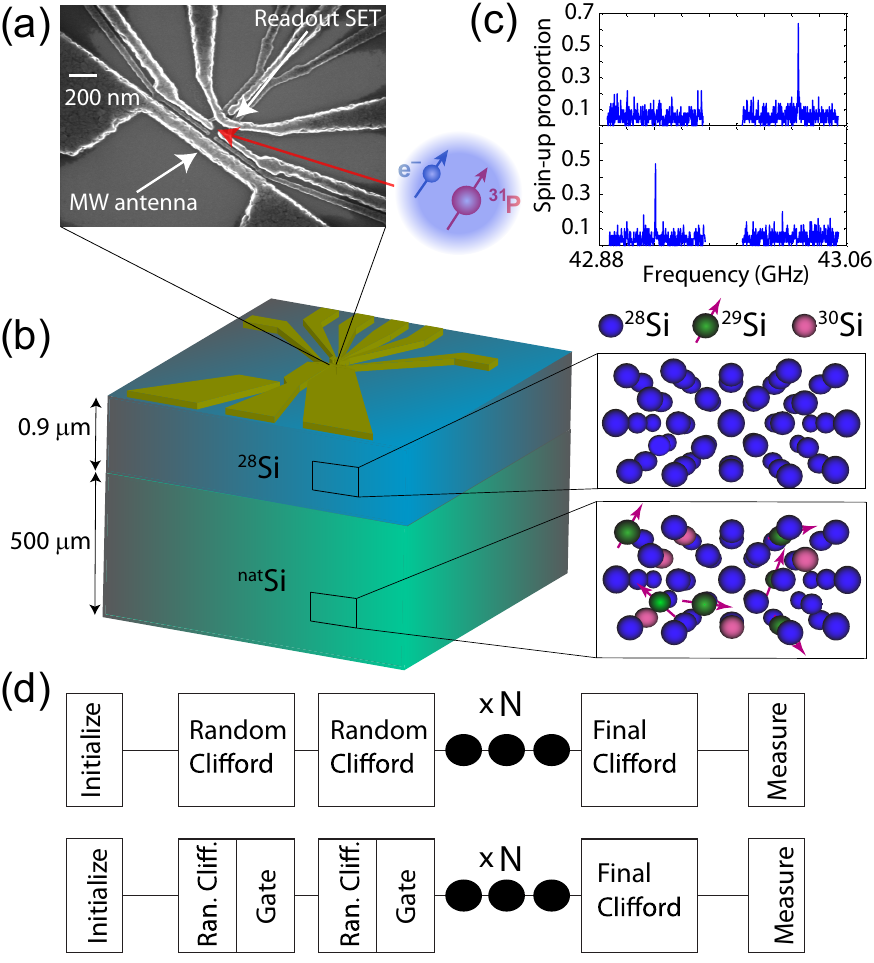}
\caption{Device structure and schematic of the benchmarking protocols. 
(a) Scanning electron micrograph image of a device similar to the one used in the experiment, highlighting the position of the P donor, the microwave (MW) antenna, and the SET for spin readout. 
(b) Schematic of the Si substrate, consisting of an isotopically purified $^{28}$Si epilayer (with a residual $^{29}$Si concentration of 800 ppm) on top of a natural Si wafer. 
(c) Electron spin resonance measurements. Each of the two resonance lines correspond to one of the possible nuclear spin orientations. Since we observe a single $^{31}$P donor, any data trace shows only one resonance.
(d) Schematic of the randomized benchmarking protocol for measuring the average Clifford gate fidelity (top) and the fidelity of individual interleaved gates (bottom).}
\label{fig:device}
\end{figure}

The device structure is shown in figure \ref{fig:device}(a). The substrate is a 0.9 $\mu$m thick epilayer of isotopically purified $^{28}$Si, grown on top of a 500 $\mu$m thick $^{nat}$Si wafer (figure \ref{fig:device}(b)). The $^{28}$Si epilayer contains 800 ppm residual $^{29}$Si isotopes. A stack of aluminium gates above the SiO$_2$ is used to induce a single-electron transistor (SET) underneath the oxide. The single-donor qubit is selected out of a small number of ion-implanted $^{31}$P atoms \cite{Jamieson2005}, placed $\approx 10$~nm below the Si/SiO$_2$ interface and underneath an additional stack of control gates. The single-shot electron spin readout is based on its spin-dependent tunneling \cite{Elzerman2004,Morello2010} into the nearby SET island. The readout process also leaves the electron spin initialized in the $\ket{\downarrow}$ state. Since this only occurs when the donor is brought in resonance with the Fermi level of the detector SET, we can tune the electrostatic gates fabricated above the donor implant area to ensure that, at any time, only a single donor is able to undergo electron tunneling events. The successful isolation of a single $^{31}$P donor is unequivocally proven by ESR experiments where only one of the ESR frequencies is active (figure \ref{fig:device}(c)) at any time. The ESR frequency sporadically jumps between $\nu_{e1}$ and $\nu_{e2}$, signalling the flip of the nuclear spin state \cite{Pla2013}. The nuclear spin readout is obtained by applying a $\pi$-pulse at frequency $\nu_{e2}$ to an electron initially $\ket{\downarrow}$. If the electron is successfully flipped to the $\ket{\uparrow}$ state, the nuclear spin state is declared $\ket{\Uparrow}$.

\section{Randomized benchmarking protocol for average Clifford gate fidelity}

The randomized benchmarking protocol we adopt is based upon the one studied theoretically in \cite{Magesan2011,Magesan2012,Epstein2014} and demonstrated experimentally with e.g. superconducting qubits and silicon quantum dots in \cite{Barends2014,Veldhorst2014}. It is based on measuring the probability of preserving an initial quantum state while applying a variable number of Clifford gate operations. Within the validity of certain assumptions \cite{Magesan2012,Epstein2014}, this probability is predicted to decay exponentially with the number of gate operations and the decay constant is related to the average Clifford gate fidelity.

Specifically, for a sequence of $N$ gate operations, we choose randomly (with uniform probability) the $N$ gates from the 24 Clifford gates, where each Clifford gate is composed of $\pi$ and $\pi/2$ rotations around different axes (table \ref{tab:Cl}). We then add a final Clifford gate to the sequence, chosen to ensure that -- if all the previous operations were ideal -- the final qubit state will be an eigenstate of the observable $\sigma_z$ accessible to our measurement. Since the qubit readout is single-shot, we have to repeat $r$ times the same gate sequence to extract the probability of recovering the expected state. For the e$^-$ qubit we used $r=200$. To average over the whole gate space, the process described above is repeated $K$ times (at constant $N$) for different random gate sets. It has been shown \cite{Epstein2014} that the measured fidelity converges to the correct average gate fidelity for $K \gg 10$. As a reasonable compromise between convergence and measurement time, we have chosen the value $K = 15$. Finally, the sequence is repeated for $n$ different values of $N$, to extract the decay of the process fidelity upon increasing $N$. A rigorous analysis of the confidence intervals for randomized benchmarking as a function of $K$ and $N$ is given in \cite{Wallman2014}.

\begin{table}
\centering
\resizebox{0.35\textwidth}{!}{
\begin{tabular}{ c |  c  }
 Clifford gate &  Physical gates  \\
 \hline
1 & I \\
2 & Y/2 \& X/2 \\
3 & -X/2 \& -Y/2  \\
4 & X \\
5 & -Y/2 \& -X/2 \\
6 & X/2 \& -Y/2  \\
7 & Y \\
8 & -Y/2 \& X/2 \\
9 & X/2 \& Y/2  \\
10 & X  \& Y \\
11 &  Y/2 \& -X/2 \\
12 & -X/2 \& Y/2  \\
 \hline
13 & Y/2 \& X \\
14 & -X/2 \\
15 & X/2 \& -Y/2 \& -X/2  \\
16 & -Y/2 \\
17 & X/2 \\
18 & X/2 \& Y/2 \& X/2  \\
19 & -Y/2 \& X \\
20 & X/2 \& Y \\
21 & X/2 \& -Y/2 \& X/2  \\
22 & Y/2 \\
23 & -X/2 \& Y \\
24 & X/2 \& Y/2 \& -X/2  \\
 \hline
 \end{tabular}
}
\caption{The Clifford group gates written as the physical gates applied. The gate numbering is arbitrary, we follow the grouping from \cite{Epstein2014}. Physical gate X denotes a $\pi$-pulse around the X-axis (i.e., unitary operation $\exp(-i\pi\sigma_x/2)$), Y/2 a $\pi/2$-pulse around the Y-axis ($\exp(-i\pi\sigma_y/4)$), etc. The average number of physical gates per Clifford gate is 1.875.}
\label{tab:Cl}
\end{table}

In order to respect the assumptions upon which the theoretical framework of Clifford-group randomized benchmarking is based, one should normally choose the final Clifford gate such that the target output state coincides with the input state ($\ket{\downarrow}$ in the case of the e$^-$ qubit )\cite{Magesan2011}. Our measurement procedure, however, benefits considerably from choosing the target output state randomly, since this allows us to exactly cancel out certain measurement errors. We then need less free parameters in our fits, which considerably improves their accuracy. Below we show that results obtained while taking a random final state do not differ quantitatively from analysing just the gate sets where the output state is $\ket{\downarrow}$. This is understandable, since the errors only accumulate significantly after $\sim$~100 gates, and the effect of one last gate that breaks the symmetry of the sequence is not expected to be important.

Physically, the electron spin readout process \cite{Morello2010} yields the probability $P_{\uparrow}$ of the qubit being $\ket{\uparrow}$ at the end of the sequence. 
Even with a perfect sequence designed to output $\ket{\uparrow}$, the measured $P_{\uparrow}$ is $< 1$ due to SPAM errors. Similarly, a perfect sequence designed to output $\ket{\downarrow}$ will yield $P_{\uparrow} > 0$. In figure \ref{fig:eRan}(a) we plot $P_{\uparrow}$ as a function of $N$ separately for sequences designed to output either $\ket{\uparrow}$ (circles) or $\ket{\downarrow}$ (squares). We define $P_{\uparrow}^{\ket{\downarrow}}$ ($P_{\uparrow}^{\ket{\uparrow}}$) to mean the measured $P_{\uparrow}$ when the target state was $\ket{\downarrow}$ ($\ket{\uparrow}$). We can fit the $\ket{\downarrow}$ data points with the function:
\begin{equation}
P_{\uparrow}^{\ket{\downarrow}}(N) = P_0^{\ket{\downarrow}} p_c^N + P_{\infty}, \label{Pdown}
\end{equation}
where $p_c$ is related to the average Clifford gate fidelity through \cite{Magesan2011}
\begin{equation}
\mathcal{F}_c = (1+p_c)/2,
\label{eq:fit}
\end{equation}
$P_0^{\ket{\downarrow}}<0$ is related to ``dark counts'' arising from SPAM errors, and $P_{ \infty}$ represents the probability of measuring $\ket{\uparrow}$ on a completely random state. If only the sequences designed to yield $\ket{\downarrow}$ had been measured, we should fit the data leaving $P_0^{\ket{\downarrow}}$, $p_c$ and $P_{\infty}$ as free fitting parameters (dashed line in figure \ref{fig:eRan}(a)). The uncertainty on $P_{\infty}$ is rather large, since it depends on the highly scattered data points at large $N$. This affects the overall confidence interval for $p_c$ and, therefore, $\mathcal{F}_c = 99.92(16)$\%. However, having measured also sequences whose target output state is $\ket{\uparrow}$, we can extract a better estimate for $P_{\infty}$ since to a good approximation it is simply given by $(P_{\uparrow}^{\ket{\uparrow}}(N=1)+P_{\uparrow}^{\ket{\downarrow}}(N=1))/2$. If we re-fit the data using a fixed value for $P_{\infty}$ as obtained above, we find a nearly identical value of $\mathcal{F}_c = 99.90(2)$\%, but now with a much improved confidence interval. This confidence interval, however, does not account for statistical errors in determining  $P_{\infty}$.

We can go one step further and fit the combined data sets $P_{\uparrow}^{\ket{\uparrow}}$ and $1 - P_{\uparrow}^{\ket{\downarrow}}$ (figure \ref{fig:eRan}(b)) with the function
\begin{equation}
\mathcal{P}(N) = \mathcal{P}_0 p_c^N + \mathcal{P}_{\infty}, \label{Ptotal}
\end{equation}
where $\mathcal{P}$ represents the probability of retrieving the correct output state, and the value of $\mathcal{P}_{\infty} = [P_{\infty} + (1 - P_{\infty})]/2 = 0.5$ is fixed by construction. This fit yields an average Clifford gate fidelity $\mathcal{F}_c = 99.90(2)$\% for the electron spin, using square pulses. In the remainder of this paper, we will use combined $P_{\uparrow}^{\ket{\uparrow}}$ and $1 - P_{\uparrow}^{\ket{\downarrow}}$ data sets to extract $\mathcal{F}_c$. The ability to fix $\mathcal{P}_{\infty} = 0.5$ greatly reduces the uncertainty in the fit to the data, but importantly does not cause  measurable deviation of the value of $\mathcal{F}_c$ from that extracted using only the sequences where the target state coincides with the initial state. We confirmed this conclusion on multiple datasets.

\section{Dependence of e$^-$ gate fidelities on pulse power and shape}

Gate errors in spin qubit systems can arise from two broad classes of physical phenomena: uncertainty in the instantaneous resonance frequency of the qubit (noise in $\sigma_z$) or imperfections in the control field power and/or duration (noise in $\sigma_x$). The errors arising from noise in $\sigma_z$ can be reduced to some extent by minimizing the duration of the control pulses (provided enough microwave power is available). This maximizes the excitation spectrum in the frequency domain, and ensures that the qubit frequency is well within the pulse spectrum. In practice, however, very short pulses can become imperfect due to the finite time resolution and bandwidth of the hardware used to gate the microwave power. This can then translate into an inaccurate calibration of the pulse length, i.e., noise in $\sigma_x$. One way to mitigate this problem is to use shaped pulses.

We have applied the randomized benchmarking protocol for the electron spin qubit with two different pulse shapes: (i) simple square pulses, and (ii) sinc-function ($\sin(x)/x$) shaped pulses. We chose a sinc function truncated at $\pm 3 \pi$ (sinc-3). For a similar length $\pi$-pulse, the sinc pulse gives an excitation spectrum that is roughly 5 times wider than that of a square pulse [figure~\ref{fig:pulses}(a)], but correspondingly requires a higher peak microwave power. At the same peak microwave power, the sinc pulse produces an excitation profile that is similar to the square pulse [figure~\ref{fig:pulses}(b)], but has a 5 times longer total duration. This can help relaxing the requirements on the time resolution of the pulse generator.

\begin{figure}
\centering
\includegraphics[width=0.45\textwidth]{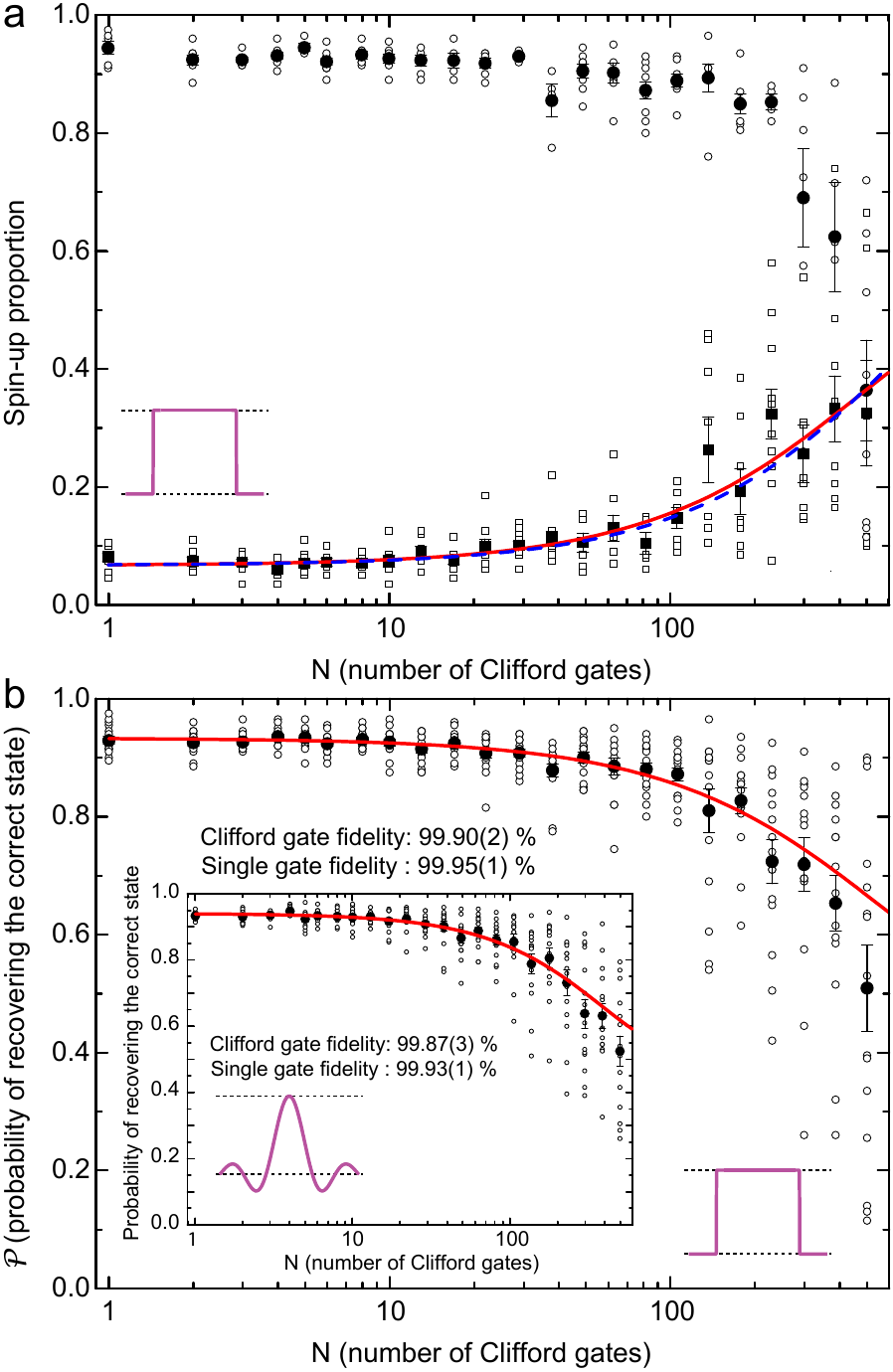}
\caption{(a) Randomized benchmarking measurements on the electron spin qubit, using square pulses at 1 mW source power ($\sim 100$~nW at the device). Data has been separated to show the case where the target output state is $\ket{\uparrow}$ (circles) or $\ket{\downarrow}$ (squares). Open symbols show the electron spin-up probability $P_{\uparrow}$ from each of the individual sequences of Clifford gates. Solid symbols show the mean and the standard error of mean for each sequence. The lines are fits of the $\ket{\downarrow}$ data to Eq.~\ref{Pdown}, with $P_{\infty}$ either as a free (dashed line) or fixed (solid line) fitting parameter. See text for details. (b) Main panel: same data as in (a), but combining the data sets for both target output states to obtain the overall probability $\mathcal{P}$ to recover the correct output state. The solid line is a fit to Eq.~\ref{Ptotal} with $\mathcal{P}_{\infty}=0.5$ as a fixed parameter. Inset: Similar data obtained with sinc pulses at 3.16 mW source power. The gate fidelity error margins are calculated from the 95~\% confidence interval for parameter $p_c$ from the non-linear least-squares fit of $\mathcal{P}(N)$. Bootstrapping the residuals of the fit gave similar results for the confidence intervals. All fits have been weighted with the inverse of the unbiased sample variance at each $N$.}
\label{fig:eRan}
\end{figure}

\begin{figure}
\centering
\includegraphics[width=0.48\textwidth]{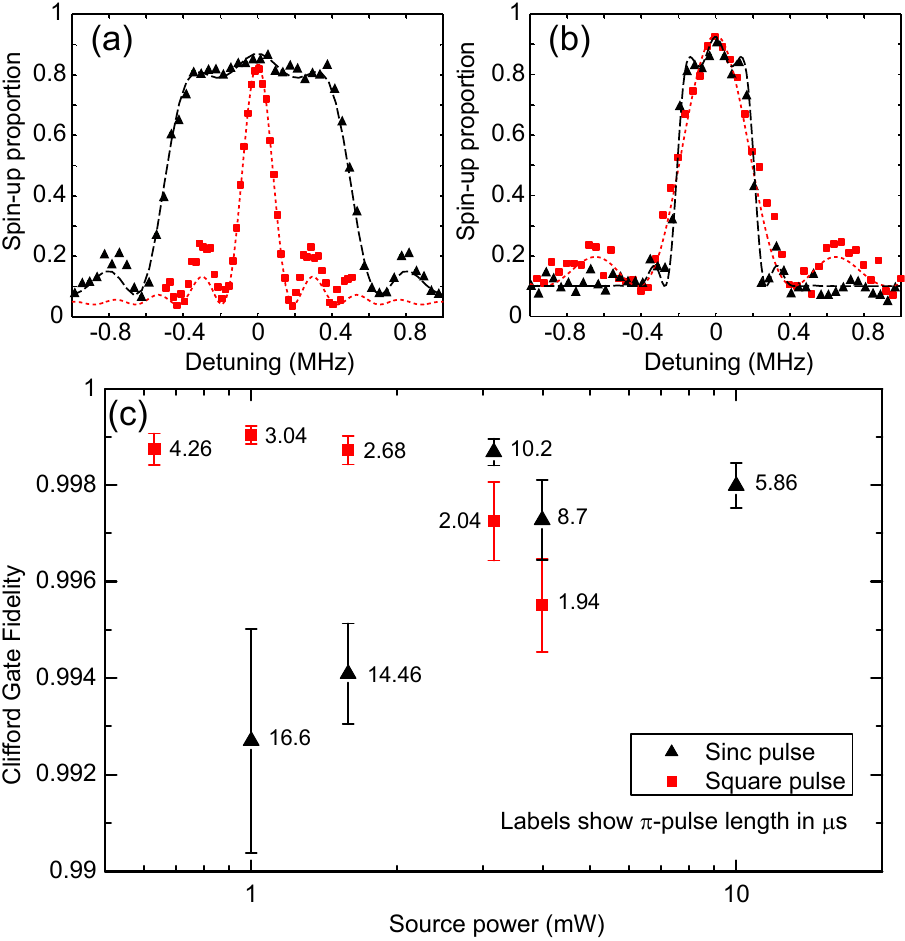}
\caption{(a-b) Measured (symbols) and simulated (lines) spin-up proportion after applying a $\pi$-pulse to a spin-down e$^-$ state as a function of pulse detuning from the resonance frequency. Triangles and dashed lines refer to sinc pulses; squares and dotted lines refer to square pulses. In (a), the microwave source power is adjusted to make the $\pi$-pulse length 4.5 $\mu$s for both pulses, whereas in (b) the peak source power is kept identical for both pulses, with $\pi$ length of 2.08 $\mu$s (11.06 $\mu$s) for a square (sinc) pulse. (c) Average Clifford gate fidelities measured as a function of source power. Labels indicate the $\pi$-pulse length in $\mu$s and triangles (squares) correspond to sinc (square) pulses. All fits have been done similarly as in figure 2(b).}
\label{fig:pulses}
\end{figure}

We also investigated the gate fidelities at different microwave powers, as explained in detail below. For the optimal power, the results are shown in the main panel of figure \ref{fig:eRan}(b) for the square pulse, and in its inset for the sinc pulse. From this data we extract an average Clifford gate fidelity of 99.90(2)~\% (99.87(3)~\% for sinc). Since a Clifford gate consists on average of 1.875 single ($\pi$ or $\pi/2$) gates (see table 1), these values can be converted to average single gate fidelities as $1-(1-\mathcal{F}_c)/1.875$. This produces 99.95(1)~\%  for the square pulse and 99.93(1)~\% for the sinc pulse. 


In figure \ref{fig:pulses}(c) we present results from the randomized benchmarking protocol as a function of the source power for the two different pulse shapes. We have also indicated the time it takes to perform a $\pi$-rotation with the corresponding power and pulse shape. From this data we can see that (i) at low powers the square pulse outperforms the sinc, and (ii) at higher powers the square pulse fidelity is markedly reduced. These effects are probably both related to the pulse length: at low powers the sinc pulse duration becomes exceedingly long, so that over the long sequence of Clifford gates some low-frequency noise begins to affect the outcome. At high powers, the square pulses become short compared to the time resolution (20~ns) of the baseband generator used to drive the digital in-phase / in-quadrature (I/Q) modulator in our vector microwave source. Below 2 $\mu$s pulse length, the 20~ns resolution of the hardware already corresponds to a 1 \% precision, which starts to be the limiting factor. Conversely, we do not see a clear improving trend while increasing the microwave power. This seems to indicate that the gate fidelities are not limited by $\sigma_z$ noise. This is consistent with the very narrow resonance linewidth ($\approx 1.8$~kHz) measured in this sample \cite{Muhonen2014}, which is indeed always much smaller than the spectral width of the control pulses.

\section{Interleaved single-gate fidelities}

In addition to the average fidelity of Clifford gates, a similar sequence has been proposed \cite{Magesan2012a} and applied \cite{Barends2014,Veldhorst2014} to extract the average fidelity of a specific individual physical gate operation. This procedure is called interleaved randomized benchmarking and is based on interleaving the gate under study within a random Clifford sequence \cite{Magesan2012a} (see figure \ref{fig:device}(c)). The random sequence of Clifford gates has the effect of randomizing the input states for the interleaved gate. Since an extra gate will normally introduce additional errors, the average fidelity of the interleaved gate can be extracted by comparing the interleaved benchmarking decay to the reference Clifford decay (figure~\ref{fig:eRan}). The fitting function is the same as for the reference decay (equation~\ref{Ptotal}, replacing $p_c$ with $p_\textrm{gate}$), and the interleaved gate fidelity is then extracted as
\begin{equation}
\mathcal{F} _\textrm{gate}= (1+p_\textrm{gate}/p_c)/2.
\end{equation}

The results of the interleaved randomized benchmarking for the electron spin qubit are shown in table I. We note that the average of the individual gate fidelities is close, although slightly lower than the average single gate fidelity deduced from the reference decay. Also notable is the similarity of the gate fidelities obtained with square and sinc pulse shapes, even though they are measured at different source powers and different pulse lengths. The $\pi$-pulses appear to give higher fidelities than the $\pi/2$-pulses. This could be due to the finite time resolution of the pulse generator, which has more effect on the shorter $\pi/2$-pulses, or it could indicate some refocussing effect of the $\pi$-pulses, which compensate for some slight qubit dephasing during the long sequences of gates.


\begin{table}
\centering
\resizebox{0.4\textwidth}{!}{
\begin{tabular}{ l | c |  c  }
\multicolumn{3}{l}{ \textbf{Electron spin}} \\
 Interleaved Gate & Square pulse & Sinc pulse \\
 \hline
X & 99.93(4) \% & 99.97(4) \% \\
Y & 100.00(5) \% & 99.95(4) \% \\
X/2 & 99.93(5) \% & 99.87(6) \% \\
Y/2 & 99.85(8) \% & 99.81(6) \% \\
-X/2 & 99.94(4) \% & 99.92(5) \% \\
-Y/2 & 99.90(5) \% & 99.81(6) \% \\
 \hline
 Average  & 99.93 \% & 99.89 \% \\
\end{tabular}
}
\caption{Results of the interleaved gate benchmarking for the electron spin qubit. Notation is the same as in table \ref{tab:Cl}. The source power used in these experiments was 0.63 mW for square pulses, and 3.16 mW with sinc pulses. The error margins are calculated by summing squarely the relative 95~\% confidence limits for parameters $p_\textrm{gate}$ and $p_c$. All fits have been done similarly as in figure 2(b).}
\label{tab:ele}
\end{table}

\section{Nuclear spin gate fidelities}

As evidenced in earlier experiments \cite{Pla2013,Saeedi2013,Muhonen2014}, the nuclear spin of the $^{31}$P donor constitutes an excellent qubit, particularly when the donor is in the ionized state. We have performed randomized benchmarking of the nuclear spin qubit, focusing on the $^{31}$P$^+$ ionized state. After the gate operations, a high-fidelity, quantum non-demolition readout of the $^{31}$P nuclear state \cite{Pla2013} can be accomplished by loading a spin-down electron and applying a $\pi$-pulse on the electron spin conditional on the nuclear spin state. The electron spin is then read out \cite{Morello2010}. In the present experiments, the electron readout is repeated 50 times to obtain the nuclear state with high fidelity. For the randomized benchmarking we use $r = 75$ repetition of each sequence of $N$ Clifford gates. At every $N$ we use $K = 5$ different, randomly chosen sequences. This low value of $K$ is sub-optimal from the point of view of converging to the average Clifford gate fidelity, but was necessary to avoid having exceedingly long experiments. These measurements already took 7 hours with $K=5$, partly due to the time needed to load the random sequences of up to $N=1000$ Clifford gates onto the baseband generator. Here we used square pulses, with 150~$\mu$s duration for a $\pi$-pulse. This means that, even for the highest $N$, the total duration of the sequence remains well below the dephasing time of the qubit, $T_{2n0}^* = 600$~ms \cite{Muhonen2014}.

\begin{figure}
\centering
\includegraphics[width=0.48\textwidth]{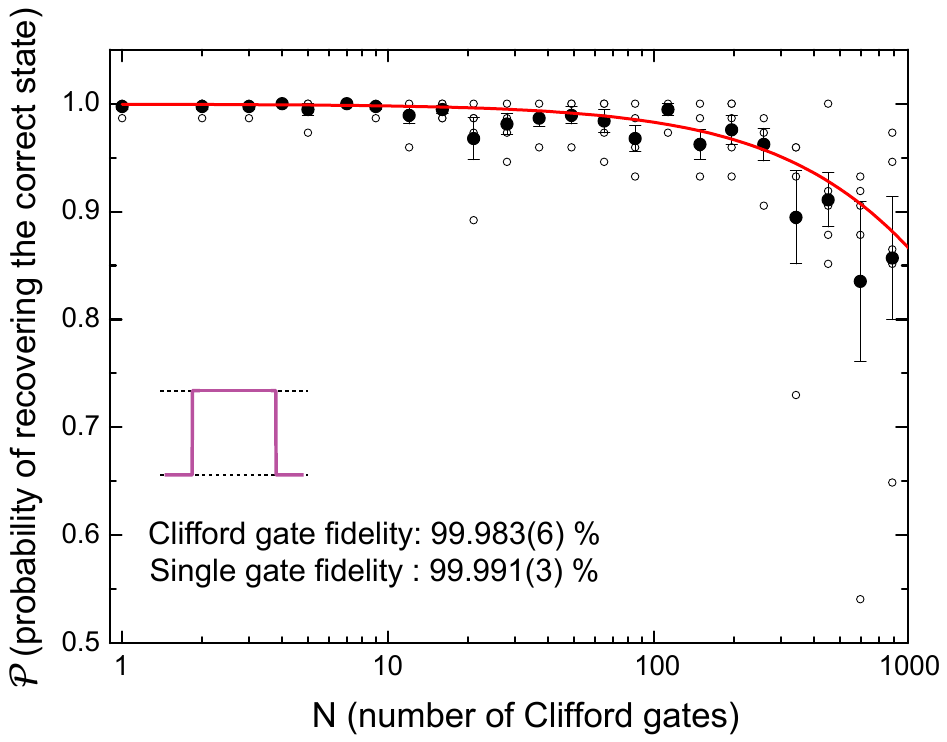}
\caption{Randomized benchmarking for the nuclear spin qubit with gate operations performed using square pulses. The duration of a $\pi$-pulse is 150~$\mu$s. The open circles show results from individual measurements, and solid circles the mean of the $K=5$ different random sequences. Error bars, fit and the error margins as in in figure 2(b).}
\label{fig:nRan}
\end{figure}

The average Clifford gate fidelity and the individual interleaved gate fidelities are shown in figure~\ref{fig:nRan} and table \ref{tab:nuc}, respectively. We found a $^{31}$P$^+$ gate fidelity of order 99.99~\%, matching even some of the stricter error correction thresholds. The individual interleaved gate fidelities are somewhat worse, but still of order 99.95\%.

\begin{table}[tb]
\centering
\resizebox{0.28\textwidth}{!}{
\begin{tabular}{ l | c  }
\multicolumn{2}{l}{ \textbf{Nuclear spin}} \\
Interleaved gate & Square pulse \\
 \hline
X & 99.94(4) \% \\
Y & 99.98(1) \% \\
X/2 & 99.97(2) \% \\
Y/2 & 99.99(1) \%  \\
-X/2 & 99.88(2) \% \\
-Y/2 & 99.93(5) \%  \\
 \hline
 Average  & 99.95 \% \\
\end{tabular}
}
\caption{Results of the interleaved gate benchmarkings for the nuclear spin qubit. All notation identical to table 2.}
\label{tab:nuc}
\end{table}

\section{Conclusions}

We have quantified the quantum gate fidelities of $^{31}$P electron and nuclear spin qubits in isotopically enriched $^{28}$Si, and showed that they exceed some fault-tolerance thresholds \cite{Knill2005,Fowler2012}. An analysis of e$^-$ gate fidelities as a function of microwave power suggests that further improvements could be obtained with higher-resolution baseband generators and other engineering solutions. The behaviour of the gate errors as a function of pulse power and shape suggests that the 1-qubit gate fidelities are not affected by device-intrinsic phenomena such as fluctuations of the qubit energy splitting. A complete assessment of the viability of donor-based spin qubits for large-scale quantum computing in the solid state will require the realization and the benchmarking of 2-qubit logic gates. A recent proposal predicts fault-tolerant 2-qubit gate fidelities with logic operations based on weak exchange interactions \cite{Kalra2014} and ESR pulses, whereas very strong exchange has been measured experimentally \cite{Dehollain2014} in a donor pair. While awaiting the experimental demonstration of a universal set of 1- and 2-qubit logic gates, the present results already show great promise for the realization of donor-spin-based quantum computers in silicon.


\section*{Acknowledgements}

We thank S. Bartlett and S. Flammia for illuminating discussions. This research was funded by the Australian Research Council Centre of Excellence for Quantum Computation and Communication Technology (project number CE11E0001027) and the US Army Research Office (W911NF-13-1-0024). We acknowledge support from the Australian National Fabrication Facility, and from the laboratory of Prof Robert Elliman at the Australian National University for the ion implantation facilities. The work at Keio has been supported in part by FIRST, the Core-to-Core Program by JSPS, and the Grant-in-Aid for Scientific Research and Project for Developing Innovation Systems	by MEXT.

\bibliographystyle{unsrt}

\begin{thebibliography}{10}

\bibitem{Shor1995}
Peter~W Shor.
\newblock Scheme for reducing decoherence in quantum computer memory.
\newblock {\em Physical review A}, 52(4):R2493, 1995.

\bibitem{Steane1996}
Andrew Steane.
\newblock Multiple-particle interference and quantum error correction.
\newblock {\em Proceedings of the Royal Society of London. Series A:
  Mathematical, Physical and Engineering Sciences}, 452(1954):2551--2577, 1996.

\bibitem{Nielsen2000}
M.~Nielsen and I.~Chuang.
\newblock {\em Quantum Computation and Quantum Information}.
\newblock Cambridge University Press, 2000.

\bibitem{Raussendorf2012}
Robert Raussendorf.
\newblock Key ideas in quantum error correction.
\newblock {\em Philosophical Transactions of the Royal Society A: Mathematical,
  Physical and Engineering Sciences}, 370(1975):4541--4565, 2012.

\bibitem{Knill1998}
Emanuel Knill, Raymond Laflamme, and Wojciech~H Zurek.
\newblock Resilient quantum computation.
\newblock {\em Science}, 279(5349):342--345, 1998.

\bibitem{Stephens2008}
AM~Stephens, AG~Fowler, and LCL Hollenberg.
\newblock Universal fault tolerant quantum computation on bilinear nearest
  neighbor arrays.
\newblock {\em Quantum Information \& Computation}, 8(3):330--344, 2008.

\bibitem{Knill2005}
E.~Knill.
\newblock Quantum computing with realistically noisy devices.
\newblock {\em Nature}, 434(7029):39--44, March 2005.

\bibitem{Fowler2012}
Austin~G. Fowler, Matteo Mariantoni, John~M. Martinis, and Andrew~N. Cleland.
\newblock Surface codes: Towards practical large-scale quantum computation.
\newblock {\em Phys. Rev. A}, 86(3):032324--, September 2012.

\bibitem{Barends2014}
R.~Barends, J.~Kelly, A.~Megrant, A.~Veitia, D.~Sank, E.~Jeffrey, T.~C. White,
  J.~Mutus, A.~G. Fowler, B.~Campbell, Y.~Chen, Z.~Chen, B.~Chiaro,
  A.~Dunsworth, C.~Neill, P.~O/'Malley, P.~Roushan, A.~Vainsencher, J.~Wenner,
  A.~N. Korotkov, A.~N. Cleland, and John~M. Martinis.
\newblock Superconducting quantum circuits at the surface code threshold for
  fault tolerance.
\newblock {\em Nature}, 508(7497):500--503, April 2014.

\bibitem{Benhelm2008}
Jan Benhelm, Gerhard Kirchmair, Christian~F Roos, and Rainer Blatt.
\newblock Towards fault-tolerant quantum computing with trapped ions.
\newblock {\em Nature Physics}, 4(6):463--466, 2008.

\bibitem{Chuang1997}
Isaac~L. Chuang and M.~A. Nielsen.
\newblock Prescription for experimental determination of the dynamics of a
  quantum black box.
\newblock {\em Journal of Modern Optics}, 44(11-12):2455--2467, November 1997.

\bibitem{Chow2009}
J.~M. Chow, J.~M. Gambetta, L.~Tornberg, Jens Koch, Lev~S. Bishop, A.~A. Houck,
  B.~R. Johnson, L.~Frunzio, S.~M. Girvin, and R.~J. Schoelkopf.
\newblock Randomized benchmarking and process tomography for gate errors in a
  solid-state qubit.
\newblock {\em Phys. Rev. Lett.}, 102(9):090502--, March 2009.

\bibitem{Emerson2005}
Joseph Emerson, Robert Alicki, and Karol Zyczkowski.
\newblock Scalable noise estimation with random unitary operators.
\newblock {\em Journal of Optics B: Quantum and Semiclassical Optics},
  7(10):S347--, 2005.

\bibitem{Knill2008}
E.~Knill, D.~Leibfried, R.~Reichle, J.~Britton, R.~B. Blakestad, J.~D. Jost,
  C.~Langer, R.~Ozeri, S.~Seidelin, and D.~J. Wineland.
\newblock Randomized benchmarking of quantum gates.
\newblock {\em Phys. Rev. A}, 77(1):012307--, January 2008.

\bibitem{Olmschenk2010}
S~Olmschenk, R~Chicireanu, K~D Nelson, and J~V Porto.
\newblock Randomized benchmarking of atomic qubits in an optical lattice.
\newblock {\em New Journal of Physics}, 12(11):113007--, 2010.

\bibitem{Brown2011}
K.~R. Brown, A.~C. Wilson, Y.~Colombe, C.~Ospelkaus, A.~M. Meier, E.~Knill,
  D.~Leibfried, and D.~J. Wineland.
\newblock Single-qubit-gate error below 10$^{-4}$ in a trapped ion.
\newblock {\em Phys. Rev. A}, 84(3):030303--, September 2011.

\bibitem{Ryan2009}
C~A Ryan, M~Laforest, and R~Laflamme.
\newblock Randomized benchmarking of single- and multi-qubit control in
  liquid-state nmr quantum information processing.
\newblock {\em New Journal of Physics}, 11(1):013034--, 2009.

\bibitem{Magesan2011}
Easwar Magesan, J.~M. Gambetta, and Joseph Emerson.
\newblock Scalable and robust randomized benchmarking of quantum processes.
\newblock {\em Phys. Rev. Lett.}, 106(18):180504--, May 2011.

\bibitem{Magesan2012}
Easwar Magesan, Jay~M. Gambetta, and Joseph Emerson.
\newblock Characterizing quantum gates via randomized benchmarking.
\newblock {\em Phys. Rev. A}, 85(4):042311--, April 2012.

\bibitem{Epstein2014}
Jeffrey~M. Epstein, Andrew~W. Cross, Easwar Magesan, and Jay~M. Gambetta.
\newblock Investigating the limits of randomized benchmarking protocols.
\newblock {\em Phys. Rev. A}, 89(6):062321--, June 2014.

\bibitem{Gottesman1998}
Daniel Gottesman.
\newblock The heisenberg representation of quantum computers.
\newblock {\em arXiv preprint quant-ph/9807006}, 1998.

\bibitem{Ager2005}
JW~Ager, JW~Beeman, WL~Hansen, EE~Haller, ID~Sharp, C~Liao, A~Yang, MLW
  Thewalt, and H~Riemann.
\newblock High-purity, isotopically enriched bulk silicon.
\newblock {\em Journal of the Electrochemical Society}, 152(6):G448--G451,
  2005.

\bibitem{Tyryshkin2012}
Alexei~M. Tyryshkin, Shinichi Tojo, John J.~L. Morton, Helge Riemann,
  Nikolai~V. Abrosimov, Peter Becker, Hans-Joachim Pohl, Thomas Schenkel,
  Michael L.~W. Thewalt, Kohei~M. Itoh, and S.~A. Lyon.
\newblock Electron spin coherence exceeding seconds in high-purity silicon.
\newblock {\em Nat. Mater.}, 11(2):143--147, February 2012.

\bibitem{Saeedi2013}
Kamyar Saeedi, Stephanie Simmons, Jeff~Z. Salvail, Phillip Dluhy, Helge
  Riemann, Nikolai~V. Abrosimov, Peter Becker, Hans-Joachim Pohl, John J.~L.
  Morton, and Mike L.~W. Thewalt.
\newblock Room-temperature quantum bit storage exceeding 39 minutes using
  ionized donors in silicon-28.
\newblock {\em Science}, 342(6160):830--833, November 2013.

\bibitem{Muhonen2014}
J.~T. {Muhonen}, J.~P. {Dehollain}, A.~{Laucht}, F.~E. {Hudson},
  T.~{Sekiguchi}, K.~M. {Itoh}, D.~N. {Jamieson}, J.~C. {McCallum}, A.~S.
  {Dzurak}, and A.~{Morello}.
\newblock {Storing quantum information for 30 seconds in a nanoelectronic
  device}.
\newblock {\em ArXiv: 1402.7140}, February 2014.

\bibitem{Morello2009}
A.~Morello, C.~C. Escott, H.~Huebl, L.~H. Willems~van Beveren, L.~C.~L.
  Hollenberg, D.~N. Jamieson, A.~S. Dzurak, and R.~G. Clark.
\newblock Architecture for high-sensitivity single-shot readout and control of
  the electron spin of individual donors in silicon.
\newblock {\em Phys. Rev. B}, 80(8):081307--, August 2009.

\bibitem{Morello2010}
Andrea Morello, Jarryd~J. Pla, Floris~A. Zwanenburg, Kok~W. Chan, Kuan~Y. Tan,
  Hans Huebl, Mikko Mottonen, Christopher~D. Nugroho, Changyi Yang, Jessica~A.
  van Donkelaar, Andrew D.~C. Alves, David~N. Jamieson, Christopher~C. Escott,
  Lloyd C.~L. Hollenberg, Robert~G. Clark, and Andrew~S. Dzurak.
\newblock Single-shot readout of an electron spin in silicon.
\newblock {\em Nature}, 467(7316):687--691, October 2010.

\bibitem{Pla2012}
Jarryd~J. Pla, Kuan~Y. Tan, Juan~P. Dehollain, Wee~H. Lim, John J.~L. Morton,
  David~N. Jamieson, Andrew~S. Dzurak, and Andrea Morello.
\newblock A single-atom electron spin qubit in silicon.
\newblock {\em Nature}, 489(7417):541--545, September 2012.

\bibitem{Pla2013}
Jarryd~J. Pla, Kuan~Y. Tan, Juan~P. Dehollain, Wee~H. Lim, John J.~L. Morton,
  Floris~A. Zwanenburg, David~N. Jamieson, Andrew~S. Dzurak, and Andrea
  Morello.
\newblock High-fidelity readout and control of a nuclear spin qubit in silicon.
\newblock {\em Nature}, 496(7445):334--338, April 2013.

\bibitem{Dehollain2013}
J~P Dehollain, J~J Pla, E~Siew, K~Y Tan, A~S Dzurak, and A~Morello.
\newblock Nanoscale broadband transmission lines for spin qubit control.
\newblock {\em Nanotechnology}, 24(1):015202, 2013.

\bibitem{Jamieson2005}
D.~N. Jamieson, C.~Yang, T.~Hopf, S.~M. Hearne, C.~I. Pakes, S.~Prawer,
  M.~Mitic, E.~Gauja, S.~E. Andresen, F.~E. Hudson, a.~S. Dzurak, and R.~G.
  Clark.
\newblock {Controlled shallow single-ion implantation in silicon using an
  active substrate for sub-20-keV ions}.
\newblock {\em Applied Physics Letters}, 86(20):202101, 2005.

\bibitem{Elzerman2004}
JM~Elzerman, R~Hanson, LH~Willems Van~Beveren, B~Witkamp, LMK Vandersypen, and
  Leo~P Kouwenhoven.
\newblock Single-shot read-out of an individual electron spin in a quantum dot.
\newblock {\em Nature}, 430(6998):431--435, 2004.

\bibitem{Veldhorst2014}
M.~{Veldhorst}, J.~C.~C. {Hwang}, C.~H. {Yang}, A.~W. {Leenstra}, B.~{de
  Ronde}, J.~P. {Dehollain}, J.~T. {Muhonen}, F.~E. {Hudson}, K.~M. {Itoh},
  A.~{Morello}, and A.~S. {Dzurak}.
\newblock {An addressable quantum dot qubit with fault-tolerant control
  fidelity}.
\newblock {\em ArXiv:1407.1950}, July 2014.

\bibitem{Wallman2014}
Joel~J Wallman and Steven~T Flammia.
\newblock Randomized benchmarking with confidence.
\newblock {\em arXiv:1404.6025}, 2014.

\bibitem{Magesan2012a}
Easwar Magesan, Jay~M. Gambetta, B.~R. Johnson, Colm~A. Ryan, Jerry~M. Chow,
  Seth~T. Merkel, Marcus~P. da~Silva, George~A. Keefe, Mary~B. Rothwell,
  Thomas~A. Ohki, Mark~B. Ketchen, and M.~Steffen.
\newblock Efficient measurement of quantum gate error by interleaved randomized
  benchmarking.
\newblock {\em Phys. Rev. Lett.}, 109(8):080505--, August 2012.

\bibitem{Kalra2014}
Rachpon Kalra, Arne Laucht, Charles~D. Hill, and Andrea Morello.
\newblock Robust two-qubit gates for donors in silicon controlled by hyperfine
  interactions.
\newblock {\em Phys. Rev. X}, 4:021044, Jun 2014.

\bibitem{Dehollain2014}
Juan~P. Dehollain, Juha~T. Muhonen, Kuan~Y. Tan, Andre Saraiva, David~N.
  Jamieson, Andrew~S. Dzurak, and Andrea Morello.
\newblock Single-shot readout and relaxation of singlet and triplet states in
  exchange-coupled $^{31}\mathrm{P}$ electron spins in silicon.
\newblock {\em Phys. Rev. Lett.}, 112:236801, Jun 2014.

\end{thebibliography}

\end{document}